\documentclass[useAMS,usenatbib]{mn2e}
\pdfoutput=1

\usepackage{graphicx}
\usepackage{amsmath,amssymb,amsfonts} 
\usepackage{multirow}
\usepackage{threeparttable}

\def\msun{$M_{\odot}$}

\def\gx339{\mbox{GX~{339-4}}}
\def\rxte{\textit{Rossi X-ray Timing Explore}}
\newcommand{\wise}{\textit{WISE}}
\newcommand{\pvalue}{\mbox{p-value}}

\title[Internal shocks driven by accretion flow variability]{Internal shocks
driven by accretion flow variability in the compact jet of the
black hole binary \gx339}
\author[S. Drappeau et al.]{S. Drappeau,$^{1,2}$\thanks{E-mail:
    samia.drappeau@irap.omp.eu} J. Malzac,$^{1,2}$ R. Belmont,$^{1,2}$
    P. Gandhi,$^{3}$ S.  Corbel,$^{4}$\\
$^{1}$Universit\'{e} de Toulouse; UPS-OMP; IRAP; Toulouse, France\\
$^{2}$CNRS; IRAP; 9 Av. colonel Roche, BP 44346, F-31028 Toulouse cedex 4,
France\\
$^{3}$School of Physics \& Astronomy, University of Southampton, Highfield, Southampton SO17 1BJ\\
$^{4}$Laboratoire AIM (CEA/IRFU - CNRS/INSU - Universit\'{e} Paris Diderot), CEA
DSM/IRFU/SAp, F-91191 Gif-sur-Yvette, France}
\begin{document}

\date{Accepted 2014 ?? ??. Received 2014 ?? ??; in original form 2014 ?? ??}

\pagerange{\pageref{firstpage}--\pageref{lastpage}} \pubyear{2014}

\maketitle

\label{firstpage}

\begin{abstract}
In recent years, compact jets have been playing a growing role in the
understanding of accreting black hole engines. In the case of X-ray binary
systems, compact jets are usually associated with the hard state phase of a
source outburst. Recent observations of \gx339\ have demonstrated the
presence of a variable synchrotron spectral break in the mid-infrared band
that was associated with its compact jet. In the model used in this
study, we assume that the jet emission is produced by electrons accelerated
in internal shocks driven by rapid fluctuations of the jet velocity. The
resulting spectral energy distribution (SED) and variability properties are
very sensitive to the Fourier power spectrum density (PSD) of the
assumed fluctuations of the jet Lorentz factor. These fluctuations are
likely to be triggered by the variability of the accretion flow which is
best traced by the X-ray emission. Taking the PSD of the jet Lorentz factor
fluctuations to be identical to the observed X-ray PSD, our study finds that
the internal shock model successfully reproduces the radio to infrared SED
of the source at the time of the observations as well as the reported strong
mid-infrared spectral variability.
\end{abstract}


\section{Introduction}
Decades after their discovery, the fine details of the mechanisms behind jet
formation and its connexion to the accretion disc are still unclear. Revealing
the disc-jet connexion would help answering major questions still open
concerning accreting black holes of all sizes, their growth and the role they
play in galaxy evolution.

Conical compact jet models have been successful in reproducing the flat, or
slightly inverted radio spectra usually seen in X-ray binary sources
\citep{Corbeletal2000, Fenderetal2000, CorbelFender2002}.
However they all require a dissipation process to compensate for the adiabatic
losses \citep{BlandfordKonigl1979}. One proposed process is the conversion of
jet kinetic energy to internal energy through internal shocks.

Internal shock jet models have been proposed to model the multi-wavelength
emission from $\gamma$-ray burst \citep{ReesMeszaros1994,
DaigneMochkovitch1998}, active galactic nuclei \citep{Rees1978a, Spadaetal2001,
BottcherDermer2010} and microquasars
\citep{KaiserSunyaevSpruit2000,JamilFenderKaiser2010, Malzac2013}.  One key
point of these models is that their resulting spectral energy distributions
(SED) are very sensitive to the shape of the assumed fluctuations of the jet
velocity \citep{Malzac2014}.  \citet{Malzac2013} has shown that internal shocks
powered by flicker noise fluctuations of the bulk Lorentz factor can entirely
compensate for the adiabatic expansion losses.  Interestingly, the X-ray power
spectrum of X-ray binaries, which traces the variability of the accretion flow
in the vicinity of the compact object, is close to a flicker noise process
\citep{Lyubarskii1997, Kingetal2004, MayerPringle2006}.

\gx339\ is a recurrent X-ray transient and is believed to harbour a black hole
in a binary system with a low-mass companion star. Although the black hole mass
and the system inclination angle, and distance are still unknown, they range
between 5.8 and 10 \msun\ \citep{Hynesetal2003, Munozdaraisetal2008,
Shidatsuetal2011}, $20^\circ$ and $50^\circ$ \citep{Milleretal2006,
DoneDiazTrigo2010, Shidatsuetal2011}, and 6 and 15 kpc \citep{Hynesetal2004,
Zdziarskietal2004, Shidatsuetal2011}, respectively. The source exhibits
multi-wavelength variability on a broad range of timescales
\citep{Motchetal1982, Fenderetal1999, Corbeletal2003, Dunnetal2008, Gandhi2009,
Casellaetal2010, Corbeletal2013}. In addition, it also shows evidence of
relativistic jets \citep{Fenderetal1997,Corbeletal2000, Markoffetal2003,
Gandhietal2008}.  The data we used in the present work are part of a
multi-wavelength study of \gx339 \citep{CadolleBeletal2011, Corbeletal2013}, and
in particular of the first mid-infrared study of the source published in
\citet{Gandhietal2011} and performed in 2010 March 11. \gx339\ was observed with
the\textit{Wide-field Infrared Survey Explorer} \citep[\wise;][]{Wrightetal2010}
satellite in 4 bands($1.36 \times 10^{13}$, $2.50 \times 10^{13}$,
$6.52 \times 10^{13}$ and $8.82 \times 10^{13}$ Hz, respectively W4, W3, W2 and
W1), at 13 epochs, sampled at multiples of the satellite orbital period of 95
minutes and with a shortest sampling interval of 11~s, when \wise\ caught the
source on two consecutive scans. Radio data were obtained with the
\textit{Australian Telescope Compact Array} (\textit{ATCA}) during two days -
closest to but not simultaneous with \wise\ data~- on 2010 March 7 and 2010
March 14. The mean fluxes are $9.1 \pm 0.1$ and $9.7 \pm 0.1$~mJy at 5.5 and 9
GHz, respectively.  X-ray data were nearly simultaneous with \wise, taken
between epochs 12 and 13 with the \rxte\ (\textit{RXTE}) satellite.
\citet{Gandhietal2011} confirms the detection in the mid-infrared of a
synchrotron break associated with the compact jet in \gx339
\citep{CorbelFender2002}, and reports the first clear detection of its strong
variability. This detection of the jet's intrinsic variability and the overall
properties of \gx339\ make it the perfect source to test our model.

The objective of this paper is to determine whether an internal shock jet model
driven by accretion flow variability reproduces spectral and timing
observations of an X-ray binary source in the hard spectral state, known to be
associated with compact jets. Our work differs from previous internal shock
jet models in that we use observed X-ray power spectral density (PSD) of the
studied source, \gx339, to constrain the fluctuations of the bulk Lorentz
factor $\gamma$ of the ejecta constituting the jet.
In Section~\ref{sec:ishmodel}, we introduce the internal shock jet model used
to perform our simulations and the assumptions chosen to model the source.
Section~\ref{sec:sed} and Section~\ref{sec:timing} present the spectral and
timing analyses carried out during this study, and the results obtained. We
conclude this paper with a discussion of these results and suggestions for
future developments in Section~\ref{sec:discussion}.

\section{Internal shock model}
\label{sec:ishmodel}
\citet{Malzac2014} presents a newly developed numerical code which simulates
the hierarchical merging and the emission of ejecta constituting a jet. In this
model, a new shell of gas is ejected at each time step $\Delta t$, comparable
to the dynamical timescale at $r_{dyn}$, the initial radius of the ejecta. The
Lorentz factor of each created shell varies, depending on the time of ejection.
Its fluctuation follows a specified PSD shape.  Throughout the duration of the
simulation, the injected shells -- and any subsequent shells resulting from
mergers -- are tracked until they interact and merge with other ejecta. The
ejecta loose internal energy via adiabatic losses when propagating outwards.
However, during mergers, a fraction of their kinetic energy is converted into
internal energy. The details of the physics and the description of the main
parameters of the model are presented in the original paper. The aim of this
study is to investigate the possibility to reproduce \gx339\ broad-band spectra
and infrared light curves measured in \citet{Gandhietal2011} with such a jet
model, using the X-ray PSD as input for the fluctuations of the bulk Lorentz
factor $\gamma$ of the jet.

Following \citet{Gandhietal2011}, we take as initial parameters representative
of the source, a mass of the central object of 10 \msun\ and a distance of 8 kpc
\citep{Zdziarskietal2004, Shidatsuetal2011}. We let our simulations run for
$t_{simu} = 10^5\,\mathrm{s}$ \mbox{($\sim 1$ day)}, to allow the jet to
develop. Due to the uncertainty on the inclination angle $\theta$, we examine
different values between 20 and 50 degrees. We set the jet opening angle $\phi$
to $1^\circ$. We simulate a counter-jet in this study. However, its
contribution to the total spectral energy distribution is less than
$10\%$ in the energy range of interest.

The total power available to the jet is an important parameter of the model. To
estimate that parameter, we follow the method described in
\citet{Kordingetal2006a} and use equation (6) and equation (8) of this paper to
relate the observed X-ray luminosity of an X-ray binary source to the power
available to its jets:
\begin{align}
    P_{jet} & \approx 1.57 \times 10^{37} \left(\frac{L_{2-10
    \mathrm{keV}}}{10^{36} \mathrm{erg} \mathrm{s}^{-1}}\right)^{0.5}\,
    \mathrm{erg/s}
    \label{equ:pjet}
\end{align}

\citet{Gandhietal2011} reports an X-ray luminosity during the observations of
$L_{2-10 \, \mathrm{keV}} = 2.0 \times 10^{37} \mathrm{erg}\, \mathrm{s}^{-1}$.
As a consequence, we estimate the total power of the jet during the observations
to be $P_{jet} \simeq 0.05 \, L_{Edd}$.

\begin{figure}
  \centering
\includegraphics[width=0.45\textwidth]{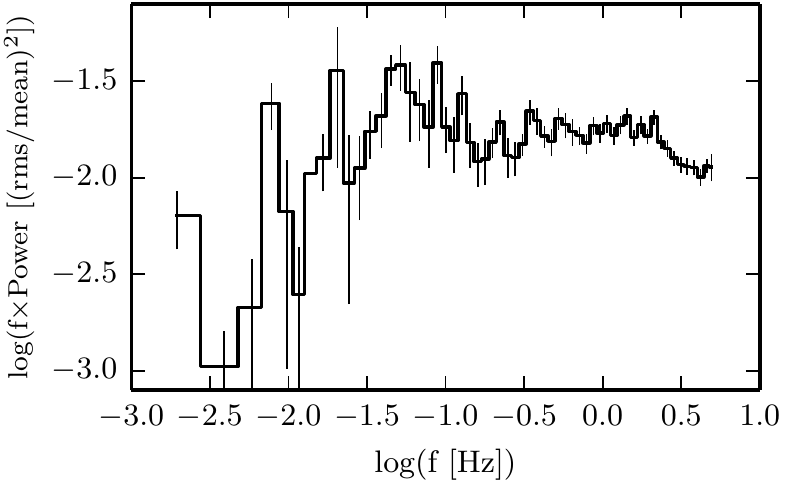}
\caption[X-ray PSD]{X-ray power spectral density (PSD) of \gx339 in the
\mbox{3-20 keV} band, used to constrain the fluctuations of the bulk Lorentz
factor of the ejecta. The PSD was extracted for the RXTE PCA observations
with ObsId 95409-01-09-03 which was quasi-simultaneous with WISE (goodtime
exposure $\sim$ 1360~s). Standard procedures were used for computing the PSD (for
details, see section 4.2 of \citet{Gandhietal2010}).}
\label{fig:psd}
\end{figure}

Finally, the most important parameter of our model is the distribution of the
fluctuations of the jet's bulk Lorentz factor. We choose for the
distribution of the fluctuations of the kinetic energy $\gamma - 1$ to follow
the shape of \gx339\ quasi-simultaneous X-ray PSD, observed by \textit{RXTE} and
shown in Fig.~\ref{fig:psd}, as X-ray PSD is thought to trace the variability of
the accretion flow in X-ray binary sources.  Moreover the fractional rms
amplitude of $\gamma - 1$ is set to be equal to that of the X-ray PSD, in this
case 35.6\%. By imposing the distribution of the fluctuations of the jet's bulk
Lorentz factor to follow the X-ray PSD, we connect the physics of the jets to
the variability of the inner part of the accretion flow.

\begin{table}
    \centering
    \caption{Parameters explored}
    \begin{tabular}{l | l}
        Parameters                      & Range of values\\
        \hline
        Inclination angle, $\theta$     & $20^{\circ}$, $30^{\circ}$, $40^{\circ}$, $45^{\circ}$, $50^{\circ}$\\
        Mean Lorentz factor, $\gamma_{mean}$   & 1.5, 2, 4\\
        Electron equipartition, $\xi_{e}$& 0.5, 1\\
        Proton equipartition, $\xi_{p}$ & 0, 0.5, 1\\
        Ejecta scheme                   & constant shell kinetic energy,\\
                                        & constant shell mass,\\
                                        & random shell mass\\
        Shock propagation scheme        & slow, fast
    \end{tabular}
    \label{tab:params}
\end{table}

In order to find the preferred set of parameters which reproduces correctly the
broad-band spectra of the source, we have investigated the following parameters
of the model: the mean jet Lorentz factor $\gamma_{mean}$, the electron and
proton equipartition factors $\xi_{e}$ and $\xi_{p}$, the ejecta scheme, and
the shock propagation scheme. $\gamma_{mean}$ sets the amplitude of the overall
spectra.  The model provides three methods to generate the ejected shells: the
ejecta have either a constant kinetic energy, or a constant mass, or their
masses randomly vary, following the same distribution as the Lorentz factor's
fluctuations. The ejecta scheme with constant shell mass provides the
most pronounced bimodal behaviour of the correlation coefficients as observed
by \citet{Gandhietal2011}. Therefore, we impose that scheme for the rest of the
study. Finally the shock propagation scheme is a parameter representing the
two treatments available in the model of the energy dissipation occurring
during a merger: a slow dissipation method, which overestimates the energy
dissipation time, and a fast dissipation method, which underestimate the
dissipation time-scale.  These parameters, and their range, are listed in
Table~\ref{tab:params}.

To assess the correctness of our models in reproducing the spectral and timing
observations of \gx339, we compared the simulated spectral energy distributions
and infrared light curves to the data. The spectral and timing analyses
performed and the results obtained are presented in the following sections.

\section{Spectral analysis}
\label{sec:sed}
We use the internal shock model to generate different scenarios of jet
formation and emission. The emission process considered in the model is solely
synchrotron self-absorbed from non-thermal electrons. The electron
distribution is a power-law of spectral index $p=2.3$, with its minimum and
maximum energies ($\gamma_{min}$ and $\gamma_{max}$) set arbitrarily and fixed
throughout the simulation. The choice of $p=2.3$ is driven empirically by the 
observed slope of the infrared spectrum interpreted as optically thin synchrotron
radiation by a power-low energy distribution of electron (see
\citealt{Gandhietal2011}). This value of $p$ is moreover consistent with typical
value expected in shock acceleration. The emission from every shell,
initially injected as well as products of mergers, is calculated. The final
SED, being compared to the data, is the time-average of all these individual
emissions over the simulation running time $t_{simu}$. The broad-band
spectra are computed from $10^{7}$ to $10^{16}\,~\mathrm{Hz}$.

It is important to note that the general shape of the simulated SED is
determined solely by the shape of the PSD used as an input to the fluctuations
of the jet Lorentz factor. The explored parameters only allow to modify the
flux normalisation or shift it in the photon frequency direction. Moreover these
parameters are degenerated, as two different sets of parameters can produce
similar spectra. Hence, reproducing the overall shape of a observed spectra
depends essentially on the shape imposed to the fluctuations of the jet's bulk
Lorentz factor.

\begin{figure*}
  \centering
  \begin{minipage}{180mm}
\includegraphics[width=0.95\textwidth]{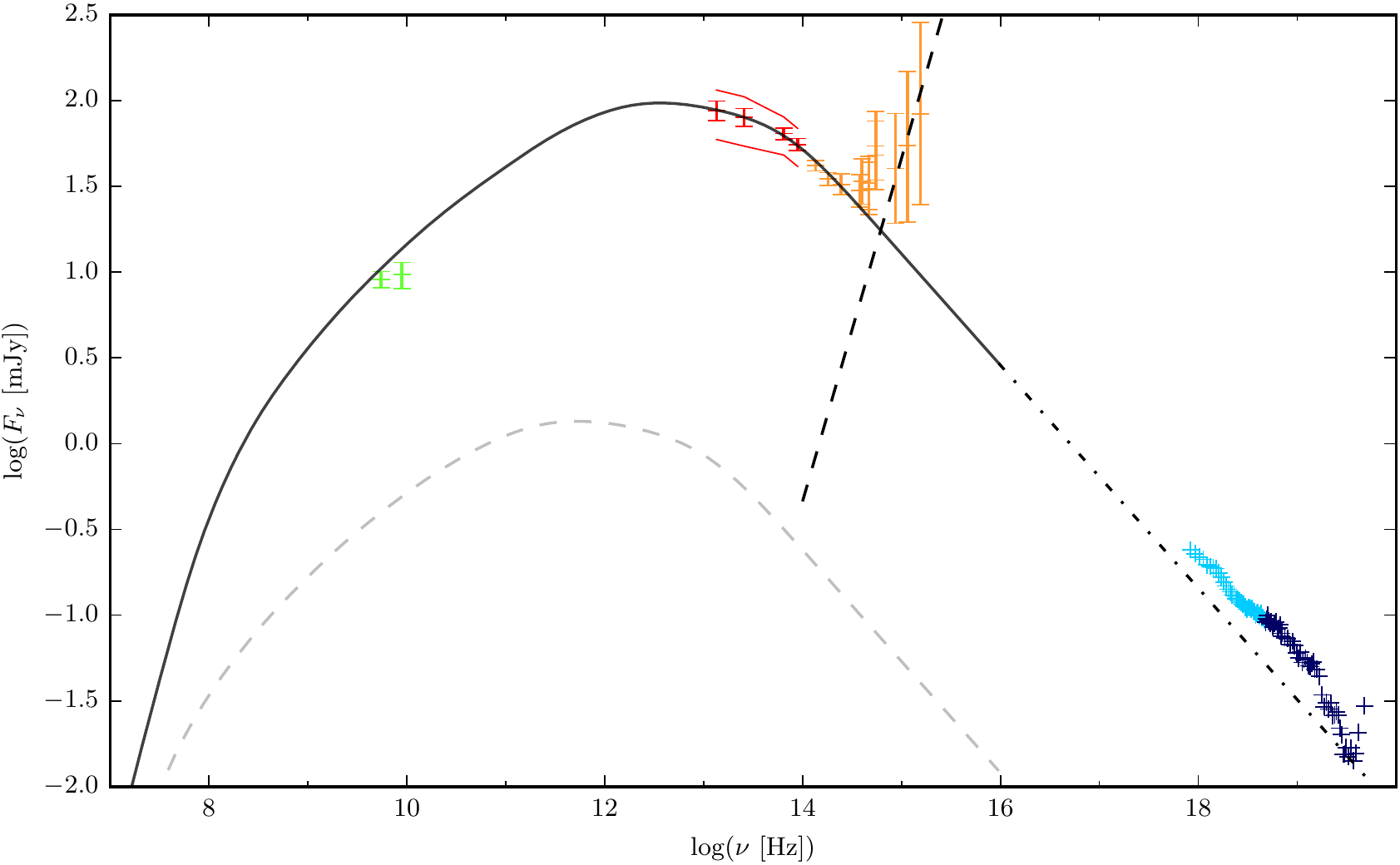}
\caption[SED of our preferred model]{Broadband spectra of our preferred model.
Data points are from \citet{Gandhietal2011}. Radio points are plotted in
green, \wise\ data in red, the near-infrared, optical and ultraviolet in
orange and the X-ray in blue. The radio points were obtained during
two days, closest to but not simultaneous with \wise\ observations. The
error bars represent the statistical and systematic errors on the mean.
The total synchrotron self-absorbed jet emission from our model is shown
as the solid black line. The contribution of the counter-jet is
represented by the dashed grey line. Red curves represent the rms amplitude
of variability over the 13 \wise\ epochs. The spectra have been averaged over
the whole duration of the simulation. The dashed and dot-dashed black lines
represent the contribution from the accretion disk to the spectra and the
extrapolation of the optically thin synchrotron jet's emission, respectively.
X-ray emission is used here solely as upper limits to define the feasibility of
our fit.}
\label{fig:preferredmodel}
\end{minipage}
\end{figure*}

Fig.~\ref{fig:preferredmodel} compares our preferred model spectral energy
distribution to the broad-band spectra of \gx339. The finding of this
preferred set of parameters is done by an approximate match of the
radio-to-infrared data. Despite the fact that no quantitative
statistical fit has been carried out on the SED, we calculate a
reduced-$\chi^2$ of 1.19 for 8 degrees of freedom (9 data points: two radio
points, four \wise\ data points and the first three optical/ultraviolet (OUV) points, and 1
free parameters: inclination angle) for this particular set of
parameters. When we do not consider the presence of a counter-jet,
the reduced-$\chi^2$ lowers to 0.80. The significant difference in the
reduced-$\chi^2$ values is mainly due to the higher contribution of the
counter-jet to the radio points ($\sim 10\%$) compared to its contribution in
the \wise\ bands ($\sim 1\%$). 5 of the 9 data points used in the calculation of
the reduced-$\chi^2$ do not include information on their variability over time.
Consequently, the reduced-$\chi^2$ does not take into account the variance of
flux in its calculation but only the error on each observed measurement. Taking
into account the flux variability of the 4 \wise\ data points would decrease
their contribution to the reduced-$\chi^2$ by roughly a factor 2 and therefore
would decrease as well the resulting reduced-$\chi^2$. It is apparent from this
figure that we are capable of reproducing the overall shape of the
radio-to-infrared observations with an internal shock model of jets powered with
the accretion flow variability of the source. One should note that the
OUV emissions are believed to originate from the outer
parts of the accretion flow.  We do not attempt to model this region. Finally,
we do not model the X-ray emission that we attribute to the accretion flow.
Instead we use the X-ray emission as an upper limit to define the feasibility of
our fit. We discuss the possible contribution of the jet to X-ray emission in
Section~\ref{sec:discussion}.

The small discrepancy between the radio data points and our model can be
partially explained by the fact that the radio observations were not performed
simultaneous with \wise\ observations. In fact, if we use only the radio flux
of the observation closest to the \wise\ observations (on 2010 March 14)
instead of the mean of the two observations, then the model lies within the
error bars of the radio points. The conical geometry assumed for the jet in
this work may have an impact on the radio emission as well.

\begin{table}
    \centering
    \caption{Preferred model parameters}
    \smallskip 
    \begin{threeparttable}
        \begin{tabular}{l | l}
            Parameters  & Values\\
            \hline
            $M_{bh}$    & 10 \msun\\
            $t_{simu}$  & $10^{5}$ s\\
            $r_{dyn}$   & 10 $r_{G}$\\
            $\Delta t$   & 9.941 ms\\
            $\phi$      & $1^{\circ}$\\
            $f_{volume}\tnote{a}$& 0.7\\
            $\gamma_{a}\tnote{b}$& 4/3\\
            $\boldsymbol{\gamma_{mean}}$& \textbf{2}\\
            PSD shape \& amplitude   & \gx339\ X-ray PSD\\
                                     & (see Fig.~\ref{fig:psd})\\
            \textbf{Ejecta scheme}& \textbf{constant shell mass}\\
            $P_{jet}$   & 0.05 $L_{Edd}$\\
            $\boldsymbol{\xi_{e}}$   & \textbf{1}\\
            $\boldsymbol{\xi_{p}}$   & \textbf{0}\\
            $p\tnote{c}$         & 2.3\\
            $\gamma_{min}$& 1\\
            $\gamma_{max}$& $10^{6}$\\
            $\boldsymbol{\theta}$    & $\boldsymbol{23^{\circ}}$\\
            \textbf{Shock propagation scheme} & \textbf{fast}
        \end{tabular}
        \begin{tablenotes}
            \item[a] volume filling factor of the colliding shells
            \item[b] effective adiabatic index of the flow 
            \item[c] spectral index of the electron distribution
        \end{tablenotes}
    \end{threeparttable}
    \label{tab:preferredmodel}
\end{table}

Table~\ref{tab:preferredmodel} presents the parameters of our preferred model.
In bold are the parameters of the model, explored in our study. These
parameters are consistent with the current knowledge we have on \gx339. However
due to the degenerate nature of the parameters, this fit is not unique.
Nevertheless, the result does provide strong hints on the physical processes
happening within the jets and its connexion to the accretion flow variability.

\section{Timing Analysis}
\label{sec:timing}
Following the finding of a preferred set of parameters, we compute the light
curves of our preferred-model on time-scale of 11 seconds, at four infrared
frequencies corresponding to the frequencies of \wise\ bands. 

One important point should be noted here. On one hand, the internal shock jet
model produces full light curves, from the beginning of the simulation to the
end. On the other hand, \wise\ satellite produces a sampling of 11-second
scans, at multiples of the satellite orbital period of 95 minutes. To enable a
comparison between the simulations and the data, one needs to apply a mask
reproducing the \wise\ scan of the source on the simulated light curves.
By doing so, we obtain a set of 13-points light curves comparable to data.

To perform the comparison of our simulated light curves to data, we examine
three characteristics: the average flux $F_{\nu}$ and the fractional
variability amplitudes $F_{var}$ of the light curves,
in each of the \wise\ infrared bands $W_{i}$, as well as the correlation
coefficients $R$ of fluxes between these bands. For a light curve consisting of
$N$ fluxes $F_{\nu}^{j}$ measured at discrete times $t_{j}$, $F_{var}$ is defined as
follows:
\begin{align}
        F_{var} & = \sqrt{\frac{S^{2}}{F_{\nu}^2}}\\
        \mathrm{with} \; S^{2} & = \frac{1}{N-1} \sum_{j=1}^{N}(F_{\nu}^{j} - F_{\nu})^{2}
\end{align}
The correlation coefficients $R$
are defined as the covariance of light curves in two bands normalised by the
product of the variances in each band.

To take into account the measurement uncertainties in these three
characteristics, we used Monte-Carlo bootstrapping to generate from
each simulated light curve a collection of noised light curves. The
noise added to each of the 13 epochs is drawn from a Gaussian
distribution with a standard deviation equal to the errors on the
individual \wise\ fluxes.

The infrared light curves are observables of a random variability process
specific to the source. The \wise\ observations provides us with one realisation
of that process. Whereas the simulations provide us with many realisations,
some of which are similar to the observations and some others are not. To
determine whether our model statistically reproduces the observations, we
investigate the distribution of simulated light curves. To that end, we use two
kinds of estimators. As a first estimator, we compute the model distribution
for each of the three characteristics $F_{\nu,W_{i}}$, $F_{var,W_{i}}$ and
$R_{W_{i}}^{W_{j}}$. 

\begin{figure}
    \centering
    \includegraphics[width=0.45\textwidth]{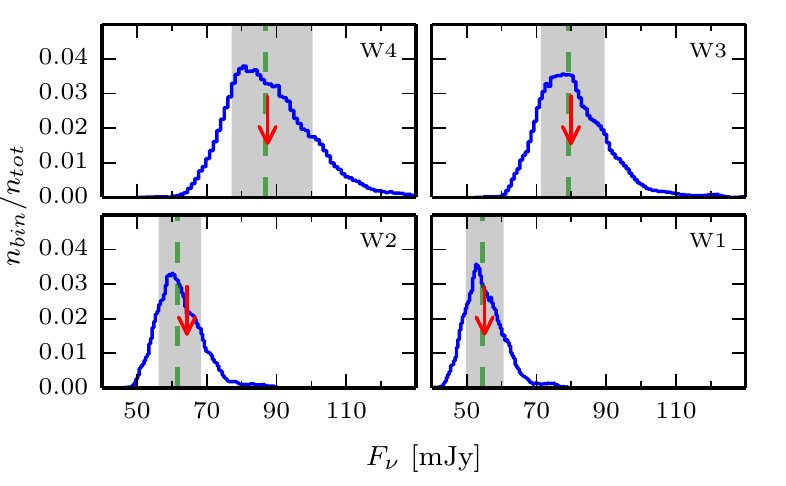}
    \caption[Distribution of $F_{\nu}$]{Distribution of $F_{\nu}$.
    The red arrow indicates the position of the observations. The green dash-line
indicates the median of the distribution, the grey area represents the
1-$\sigma$ variation about the median.}
    \label{fig:flx}
\end{figure}

\begin{figure}
    \centering
    \includegraphics[width=0.45\textwidth]{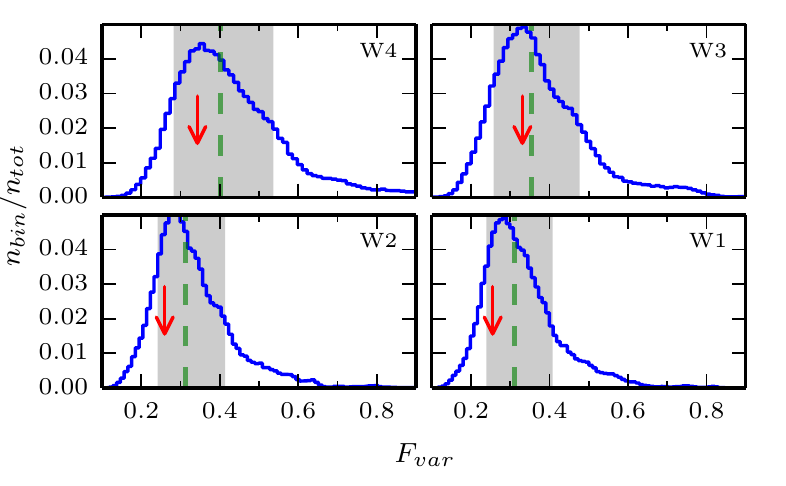}
    \caption[Distribution of $F_{var}$]{Distribution of $F_{var}$.
    The red arrow indicates the position of the observations. The green dash-line
indicates the median of the distribution, the grey area represents the
1-$\sigma$ variation about the median.}
    \label{fig:fvar}
\end{figure}

\begin{figure}
    \centering
    \includegraphics[width=0.45\textwidth]{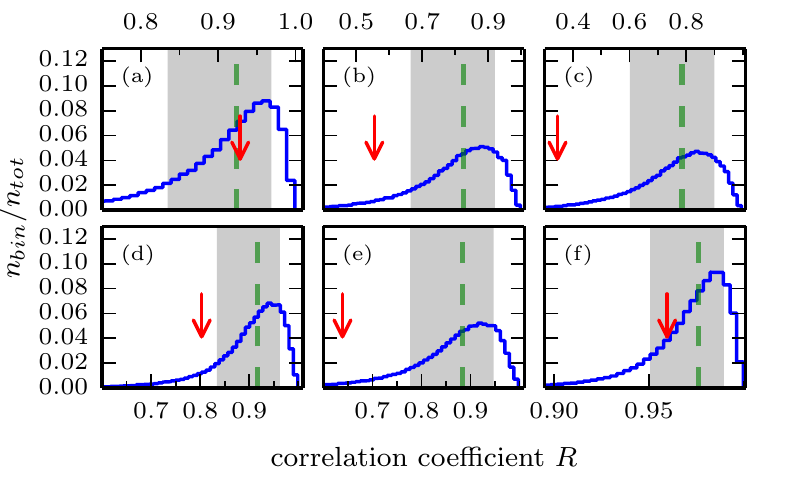}
    \caption[Correlation coefficient distributions]{Correlation coefficient distributions
        of the simulated light curves of our preferred model. Comparing: (a)
        W4-W3, (b) W4-W2, (c) W4-W1, (d) W3-W2, (e) W3-W1, and (f) W2-W1. The
        red arrow indicates the position of the observations. The green dash-line
    indicates the median of the distribution, the grey area represents the
1-$\sigma$ variation about the median.}
    \label{fig:rfactor}
\end{figure}

Fig.~\ref{fig:flx}, Fig.~\ref{fig:fvar} and Fig.~\ref{fig:rfactor} present the
distributions of $F_{\nu}$, $F_{var}$ and $R$, respectively, of all the noised
simulated light curves of our preferred model. Indicated on the figures by a
red arrow is the position of the \wise\ observations within each distribution.
In addition, the figure shows in grey area the 68\% variation around the
median of the distribution, corresponding to a `1-$\sigma$' variation
if the distributions were Gaussian. To investigate if the \wise\
observations are typical in the framework of our model (our
null-hypothesis), we evaluate the \pvalue\ of each observation
characteristic. The \pvalue\ is the probability of obtaining a test statistic
result at least as extreme as the one that was actually observed, assuming that
the null hypothesis is true. For the flux and the fractional variability
amplitude, the \pvalue s are greater than 0.1 and we cannot reject our
null-hypothesis. However, for the correlation coefficients of W4-W2,
W4-W1 and W3-W1, the \pvalue s are between 0.01 and 0.05. It suggests
that our model does not fully reproduce the correlations seen between
the bands with \wise. Indeed, the simulated light curves prove to be
more deeply correlated than the observations. Nevertheless, it is
interesting to note that the evolution of the mean values of the
correlation coefficients $R$ in Fig.~\ref{fig:rfactor} follows that
of the observations.  In the model, similar to the observations, bands W4
and W3, as well as W2 and W1 are the most correlated. Whereas W4 and W1
and W4 and W2 are the least correlated. Suggestions to explain the discrepancy
in the correlation coefficients between the model and the observations are
discussed in Section~\ref{sec:discussion}.

\begin{figure*}
  \centering
  \begin{minipage}{180mm}
\includegraphics[width=0.95\textwidth]{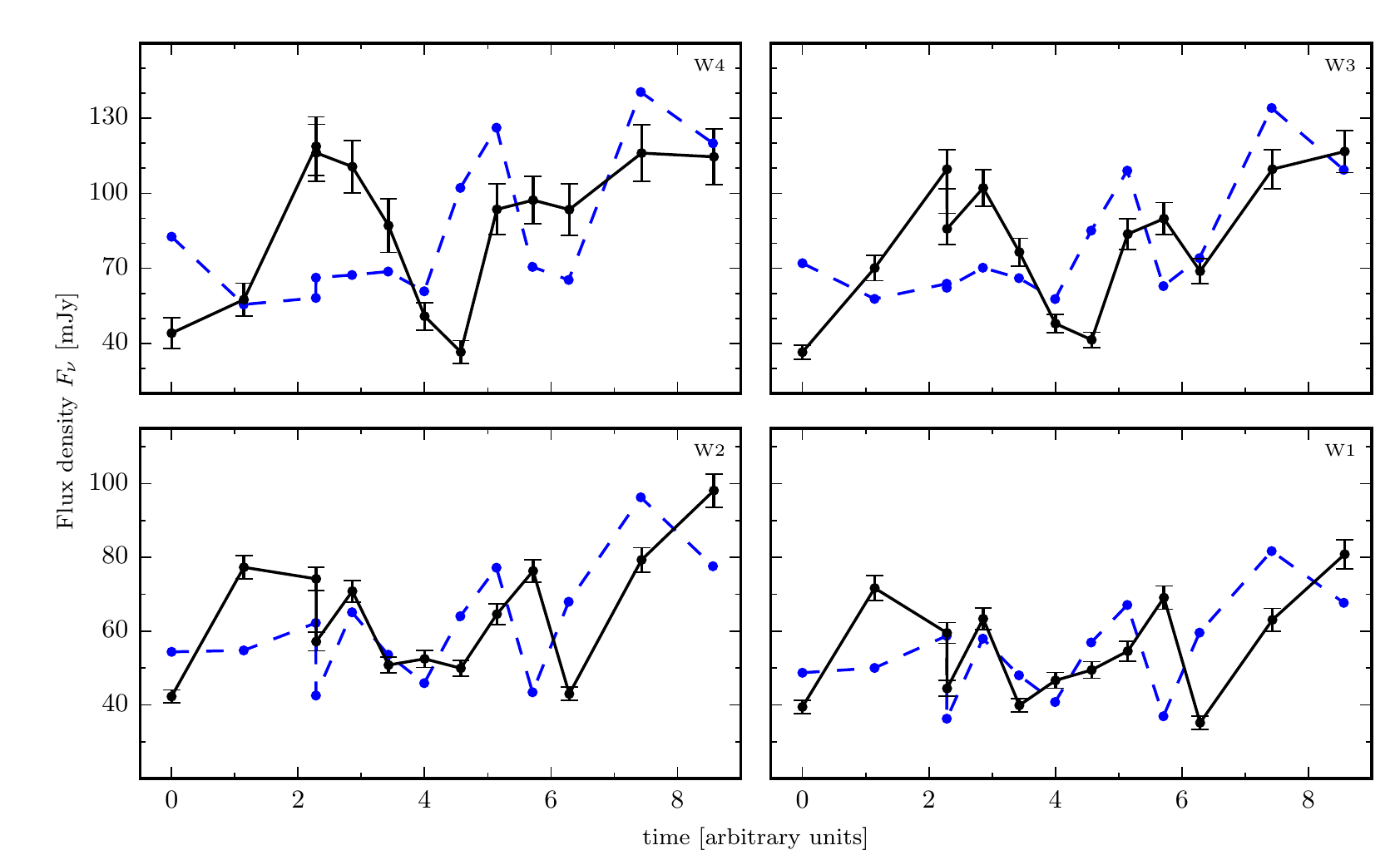}
\caption[Light curves reproducing the observations] {Selected set of light
curves reproducing the observations: in blue dash, the simulated light
curves, and in black solid, \wise\ light curves from \citet{Gandhietal2011}.
The error bars represent the statistical and systematic errors.}
\label{fig:lc}
\end{minipage}
\end{figure*}

Fig.~\ref{fig:lc} presents a particular set of simulated light curves from our
preferred model which have spectral as well as timing properties comparable to
the \wise\ observations. That is, the simulated light curves shown here are the
ones presenting simultaneously fluxes, variability and correlation coefficients
closest to the red arrows indicating in Fig .~\ref{fig:flx},
Fig.~\ref{fig:fvar} and Fig.~\ref{fig:rfactor}. Table~\ref{tab:stats}
shows the values obtained for $F_{\nu}$, $F_{var}$ and $R$ of that particular
set of simulated light curves and compare them to those reported by
\citet{Gandhietal2011}. In addition, the table reports the exact p-values
obtained for each \wise\ characteristic.

\begin{table}
    \centering
    \caption{Statistics of our preferred set of simulated light curves compared
    to those of the \wise\ observations.}
    \begin{tabular}{c | l | l | l | l}
                            &       & simulation    &   \wise           & \pvalue\\
            \hline                                  
            $F_{\nu}$ [mJy] & W1    & 54.63         & 55.2 $\pm$ 3.9    &  0.45\\
                            & W2    & 61.98         & 64.3 $\pm$ 4.6    &  0.33\\
                            & W3    & 78.81         & 79.9 $\pm$ 7.3    &  0.47\\
                            & W4    & 83.41         & 87.4 $\pm$ 8.3    &  0.47\\
            \hline                                  
            $F_{var}$       & W1    & 0.24          & 0.25 $\pm$ 0.03   &  0.23\\
                            & W2    & 0.25          & 0.25 $\pm$ 0.03   &  0.24\\
                            & W3    & 0.30          & 0.32 $\pm$ 0.06   &  0.44\\
                            & W4    & 0.35          & 0.32 $\pm$ 0.06   &  0.36\\
            \hline                                  
            R               & W4-W3 & 0.97          & 0.93              &  0.47\\
                            & W4-W2 & 0.81          & 0.55              &  0.07\\
                            & W4-W1 & 0.76          & 0.35              &  0.03\\
                            & W3-W2 & 0.91          & 0.80              &  0.10\\
                            & W3-W1 & 0.87          & 0.64              &  0.04\\
                            & W2-W1 & 0.99          & 0.96              &  0.24\\
    \end{tabular}
    \label{tab:stats}
\end{table}

The distributions of produced light curves indicate that, taken separately, the
variance of each characteristic of our model is reasonably consistent with the
observations. However, $F_{\nu}$, $F_{var}$ and $R$ are correlated
variables. Investigating them separately may bias our results. To
overcome this bias, we define, as the second estimator, a quantity
which combines all three characteristics, for each noised simulated light curve
$s$ and for the observations, as follows:

\begin{align}
    \frac{\chi^{2}_{s}}{N} = & \left(\mathbf{X}_{s} - <\mathbf{X}_{s}>\right) \cdot \left(\mathbf{Q} \otimes \left(\mathbf{X}_{s} - <\mathbf{X}_{s}>\right)\right)\\
    \frac{\chi^{2}_{obs}}{N} = & \left(\mathbf{X}_{obs} - <\mathbf{X}_{s}>\right) \cdot \left(\mathbf{Q} \otimes \left(\mathbf{X}_{obs} - <\mathbf{X}_{s}>\right)\right)
\end{align}

\noindent
where N = 8 or 14 is the number of degrees of freedom, $\mathbf{X}_{s}$ and
$\mathbf{X}_{obs}$ are matrices representing the characteristics -
$F_{\nu}$, $F_{var}$ and $R$ - in all four of \wise\ energy bands for
each noised simulated light curves~$s$ and for the observations
respectively, $<\mathbf{X}_{s}>$ is the mean of $\mathbf{X}_{s}$ over $s$, and $\mathbf{Q}$ the
inverse of the covariance matrix.

This quantity is similar to a reduced-$\chi^{2}$, with one important difference:
the distribution of the characteristics are not Gaussian.The $\chi^2$
describes the match between the modelled and observed values of the three
characteristics, exactly as it describes the match between the flux at different
energies in usual spectral fitting for instance. As the model is intrinsically
stochastic, we do not aim at reproducing accurately each single observation.
Instead we measure deviation of the observations to the model by comparing the
statistical properties of those.  As these characteristics are likely to be
correlated, we use the inverse $\mathbf{Q}$ of the covariance matrix to compute the
$\chi^2$ rather than using the simpler expression for independent variables. 
 
\begin{figure}
  \centering
\includegraphics[width=0.45\textwidth]{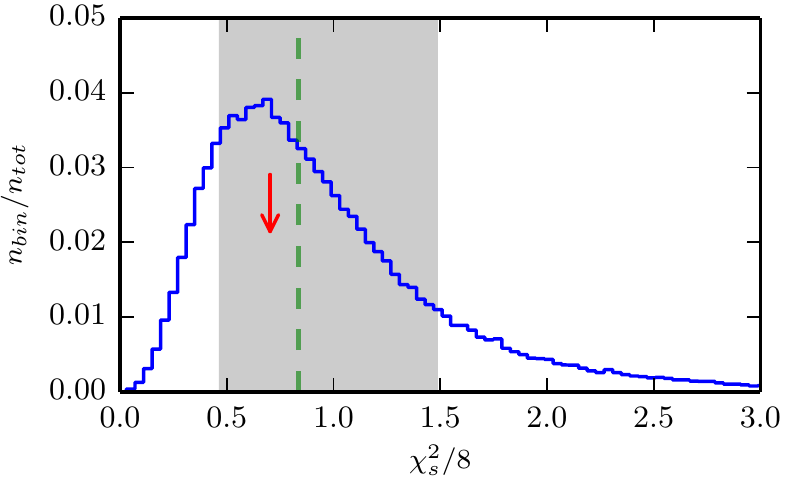}
\includegraphics[width=0.45\textwidth]{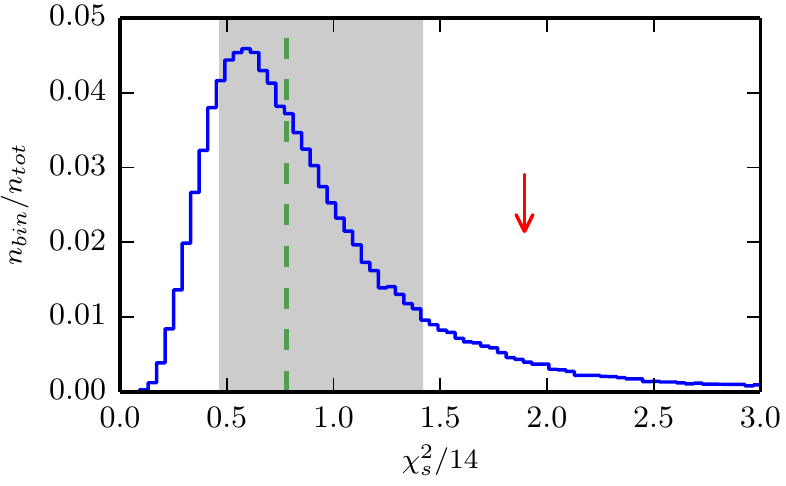}
\caption[Distribution of distances] {Distributions of reduced-$\chi^{2}_{s}$:
(top) considering $F_{\nu}$ and $F_{var}$ only; (bottom) considering
$F_{\nu}$, $F_{var}$ and $R$. The red arrow indicates the position of
the corresponding reduced-$\chi^{2}_{obs}$. The green dash-line indicates the median
of the distribution, the grey area represents the 1-$\sigma$ variation
about the median.}
\label{fig:distance}
\end{figure}

Fig.~\ref{fig:distance} presents two distributions of that quantity. On one
hand, the reduced-$\chi^{2}$ was calculated considering the average flux and
the fractional variability amplitudes only (top panel). On the other hand, it
was calculated also including the correlation coefficients (bottom panel). The
reduced-$\chi^{2}$ of the observations (represented by the red arrow) is equal
to 0.71 and 1.89, respectively. The discrepancy between the model and the
observations regarding the correlation coefficients reflects in the relatively
high reduced-$\chi^{2}_{obs}$ obtained when considering all three
characteristics in its calculation. Alternatively, the lower
reduced-$\chi^{2}_{obs}$, obtained when considering $F_{\nu}$ and $F_{var}$
only, suggests that our model correctly reproduces the spectral and timing
properties of the source at the time of the observations but does not reproduce
the correlated nature of the bands. The p-values obtained for each case
(0.38 and 0.08, respectively) suggest a similar conclusion.

\section{Discussion}
\label{sec:discussion}
The present work was designed to study the connexion between the inner part of
an accretion flow and the base of the corresponding launched outflow, in X-ray
binary sources in the hard state. Using \wise\ observations of \gx339\ on 2010
March 11 as a test-case, the results of this investigation show that it is
indeed possible to reproduce simultaneously broad-band spectral and timing
observations of a X-ray binary source in the hard state with an internal shock
jet model. The required condition is to use the X-ray timing information
provided by the corresponding power spectral density as an input to the
fluctuations of the bulk Lorentz factor $\gamma$. This finding corroborates the
ideas of \citet{JamilFenderKaiser2010} and \citet{Malzac2014}, who suggested
that accretion flow variability, traced by X-ray timing information, could
drive internal shock in jets.

Another point that could be investigated to characterise the disc-jet connexion
is the jet's contribution to the X-ray variability. Emission produced by the
model at frequencies higher than $10^{16}$ Hz are currently extrapolated from
the optically thin synchrotron power-law tail. However several cooling
processes, such as synchrotron self-Compton or inverse-Compton of the disc
emission for example, are not considered in the model yet. Moreover, the form
as well as the maximum energy of the emitting particle distribution are
currently being fixed and arbitrarily set throughout the simulation. These
restrictions lead to the choice of neglecting the possible contribution of the
jet to the X-ray variability in the present state of the model.

\begin{figure}
    \centering
    \includegraphics[width=0.45\textwidth]{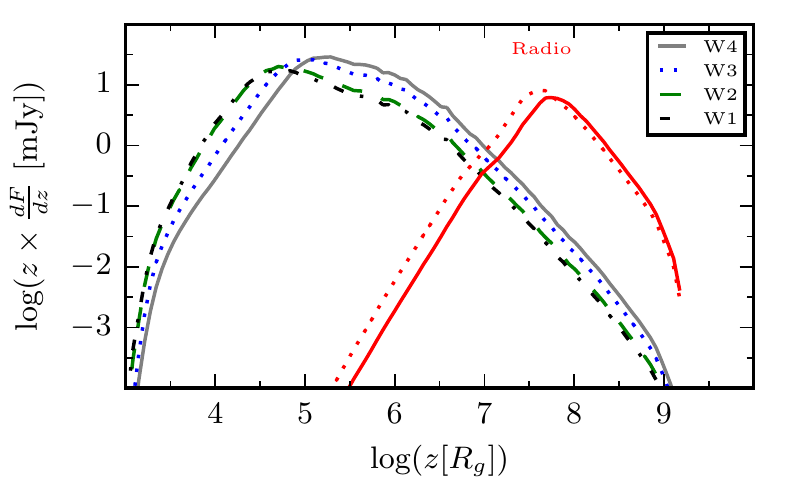}
    \caption[Regions of emission of IR and radio]{Regions of emission
                    in the infrared and radio bands. In red, are the radio
                    emission at 5.5 GHz (solid) and 9 GHz (dot). In solid
                    black, dot blue, dash green and dot-dash black are the
    infrared emission in W4, W3, W2 and W1 bands, respectively.}
    \label{fig:maps}
\end{figure}

Despite the more pronounced correlation in the simulated light curves
compared to the observations, it is interesting to note the evolution of
the spreading of the correlation coefficients $R$ in
Fig.~\ref{fig:rfactor} follows that of the observations. This
tendency may be explained in terms of regions of emission.
Fig.~\ref{fig:maps} illustrates this trend by showing that the regions
of peak emission of W1 and W2 on one hand, and W3 and W4 on the other
hand, are closely located, while the regions of peak emission of W1 and
W4 are the most distant to each other. We may improve the correspondence
between the correlation coefficients of the model and the observations by
increasing the contribution of the counter-jet to the overall emission. This is
done by releasing the constraint on the value of $P_{jet}$ and increasing the
inclination angle to $50^{\circ}$. We find a new fit to the SED by setting
$P_{jet}$ to $0.14 \, L_{Edd}$. However that new choice of parameters does not
enhance the overall fit (new reduced-$\chi^2$ is equal to 4.17) and ameliorates only
slightly the correspondence between the correlation coefficients of the model
and the observations.

The stronger correlation coefficients seen in the model, compared to
those of the observation, is due to the regions of peak emission being so
close to each others that a fluctuation occurring in one band does not have
the time to be totally dissipated before arriving in the emitting region of
the next band. To reduce the correlation between bands, one would
need to allow the particles to cool on a faster timescale or over a longer
period of time. Improving the modelling of the particle distribution by taking
into account radiative processes or modifying the geometry of the jet to
consider amplified radial expansion at the shock lead by an increase of the
internal pressure and possible radial contractions in between shock regions may
adapt the cooling timescale to that needed. Similarly, modifying the scale of
the dissipation profile along the jet by changing the one-to-one relation
between the accretion flow variability and the ejecta velocity distribution used
in this study may provide a solution to that issue.

Nevertheless, it is important to note that this disagreement between the
observed and modelled correlation coefficients does not challenge internal
shocks as a dissipation process occurring in astrophysical jets. Such
disagreement would appear in any model using any other dissipation process, as
long as they also assume a jet of conical geometry and similar radiative
treatment. Overall, investigating the variable properties of our model,
improving the modelling of the particle distribution or modifying the geometry
of the jet will shed some light on the origin of that dichotomy and will help to
understand even better the deep relation between accretion and ejection
processes in accreting black holes.

To conclude, this work is a step forward toward revealing the details of the
accretion-ejection process in accreting black hole systems. The results of this
study indicate that the conversion of jet kinetic energy to internal energy
through internal shocks may well be the dissipation process needed to compensate
for the adiabatic losses in conical compact jets. Furthermore, the
results of this work support the idea of the importance of the X-ray
variability on jet emission strength. A recent study by \citet{Dinceretal2014}
corroborates this idea. They observed that the standard
sources in the radio/\mbox{X-ray} luminosity relation show stronger
broadband \mbox{X-ray} variability than outliers at a given \mbox{X-ray}
luminosity. Further radio, mid-infrared and X-ray timing observations will
provides better constraints to the accretion-ejection connexion and to the
present model. In addition, understanding and being able to reproduce the
spectral, timing and correlation properties of a source will allow the model to
provide useful predictions to future observations, with information such as at
which frequency one can expect to observe maximum jet variability or at which
timescale is the best for probing it.

\section*{Acknowledgements}
J.M. thanks the Institute of Astronomy (Cambridge) for hospitality. P.G.
acknowledges support from STFC (grant reference ST/J003697/1). The authors are
grateful to the anonymous referee for his/her helpful comments on the
manuscript. This work is part of the CHAOS project ANR-12-BS05-0009 supported by
the French Research National Agency (http://www.chaos-project.fr).

\bibliographystyle{mn2eMOD}
\bibliography{/Users/samiadrappeau/Research/bibliography/references}{}

\bsp

\label{lastpage}

\end{document}